\documentclass[twocolumn,showpacs,preprintnumbers,amsmath,amssymb]{revtex4}
\usepackage{latexsym}
\usepackage{graphicx}% Include figure files
\usepackage{dcolumn}% Align table columns on decimal point
\usepackage{bm}% bold math
\usepackage{xcolor}
\usepackage{float}

\input amssym.def
\input amssym.tex

\newcommand{\half}{{{\textstyle\frac{1}{2}}}}

\newcommand{\be}{\begin{equation}}
\newcommand{\ee}{\end{equation} }
\newcommand{\beqa}{\begin{eqnarray} }
\newcommand{\eeqa}{\end{eqnarray} }
\newcommand{\ba}{\begin{array}}
\newcommand{\ea}{\end{array}}

\newcommand{\SO}{\mathbf{SO}}
\newcommand{\GL}{\mathbf{GL}}

\newcommand{\ODD}{\mathbf{O}(D,D)}
\newcommand{\soDD}{\mathbf{so}(D,D)}

\newcommand{\etaodd}{{\cJ}}
\newcommand{\SOD}{{\SO(1,D{-1})}}
\newcommand{\oSOD}{{\overline{\SO}(1,D{-1})}}

\newcommand{\mbg}{\mathbf{g}}

\newcommand{\G}{\mathbf{G}}
\newcommand{\DO}{\mathbf{\nabla}}
\newcommand{\doD}{\hat{D}}

\newcommand{\trd}{{\bigtriangledown}}

\newcommand\Tr{{\rm Tr}}

\newcommand\rd{{\rm d}}

\newcommand\eff{{\rm eff.}}

\newcommand\cC{{\cal C}}
\newcommand\cD{{\cal D}}

\newcommand\cF{{\cal F}}
\newcommand\cG{{\cal G}}
\newcommand\cH{{\cal H}}

\newcommand\cJ{{\cal J}}

\newcommand\cM{{\cal M}}

\newcommand\cP{{\cal P}}

\newcommand\cR{{\cal R}}

\newcommand\bcP{{\bar{\cP}}}

\newcommand\hcF{{\hat{\cal F}}}
\newcommand\hcL{{\hat{\cal L}}}

\newcommand\hdelta{{\hat{\delta}}}

\newcommand\gYM{g_{\rm \scriptscriptstyle{YM}}}

\newcommand\dis{\displaystyle}

\def\tx{\tilde{x}}

\def\bre{\bar{e}}
\def\breta{\bar{\eta}}

\def\brk{{\bar{k}}}
\def\brl{{\bar{l}}}
\def\brm{{\bar{m}}}
\def\brn{{\bar{n}}}
\def\brp{{\bar{p}}}
\def\brq{{\bar{q}}}

\def\bromega{{\bar{\omega}}}
\def\brOmega{{\bar{\Omega}}}

\def\brB{\bar{B}}
\def\brH{\bar{H}}

\def\brU{{\bar{U}}}
\def\brV{{\bar{V}}}
\def\brP{{\bar{P}}}

\def\Tw{{T}}

\begin{document}

\title{Stringy differential geometry, beyond Riemann}

\author{Imtak Jeon${}^{\dagger}$,  Kanghoon Lee${}^{\natural}$     
and Jeong-Hyuck Park${}^{\dagger}$}

\affiliation{{~}\\
\mbox{${}^{\dagger}$Department of Physics,  Sogang University, Seoul 121-742, Korea}\\
\mbox{${}^{\natural}$Center for Quantum Spacetime,  Sogang University, Seoul 121-742, Korea}\\
~\\
\textbf{\rm\small{imtak@sogang.ac.kr,~~kanghoon@sogang.ac.kr,~~park@sogang.ac.kr}}}
%${}^{\dagger}${\rm\scriptsize{Corresponding electronic address}: park@sogang.ac.kr}}

%\date{\today}

\begin{abstract}
While  the fundamental object in Riemannian geometry is a metric, closed string theories call for us to put  a  two-form  gauge field and a scalar  dilaton  on an equal footing with  the metric. Here we propose a novel differential geometry which treats the  three objects in a unified manner,  manifests not only  diffeomorphism and  one-form gauge symmetry  but also  $\ODD$ T-duality, and enables us to  rewrite  the known low energy effective action of them as a single term.  Further, we develop a corresponding vielbein formalism and gauge the internal symmetry  which is given by   a direct product of two local  Lorentz groups, ${\SOD\times\oSOD}$. We comment that   the notion of cosmological constant naturally  changes.
%%%
%%Further, we develop a corresponding vielbein formalism  and  identify  a rank four-tensor  which may provide  powerful toolkits  to organize  the higher order  derivative corrections to the stringy effective action. We comment  that the notion of cosmological constant naturally  changes.
\end{abstract}

%%%
%%02.40.-k 	Geometry, differential geometry, and topology
%%11.25.-w 	Strings and branes
%%04.60.Cf 	Gravitational aspects of string theory
%%%
\maketitle

%%%%%%%%%%%%%%%%%%%%%%%%%%%%%%%%%%%%%%%%%%%%%%%%%%%%%%%%%%%%%%%%%%%%%%%%%%%%%%%%%%%%%%%%%%%%%%%%
%%%%%%%%%%%%%%%%%%%%%%%%%%%%%%%%%%%%%%%%%%%%%%%%%%%%%%%%%%%%%%%%%%%%%%%%%%%%%%%%%%%%%%%%%%%%%%%%
\section{Introduction} 
Symmetry  guides   the structure of  Lagrangians and organizes  the physical laws into  simple forms. For example,  in Maxwell theory, the Abelian gauge symmetry  does not allow for an explicit  mass term of the vector potential, and   Lorentz symmetry unifies   the original Maxwell's four equations into two. \\
\indent In general relativity,  where the key quantity  is the spacetime metric, the diffeomorphism symmetry  first  demands  replacing   ordinary derivatives by  covariant derivatives which involve  a connection.  Setting  the metric to be   covariant  constant determines   the (torsionless) connection, \textit{i.e.~}the Christoffel symbol,   in terms of the metric and its derivatives, and hence  diffeomorphism uniquely picks up the scalar curvature as the covariant term which is   lowest order in derivatives  of the metric.  \\
\indent On the other hand, in string theories, the metric,  $g_{\mu\nu}$,  accompanies  a   two-form gauge field, $B_{\mu\nu}$,  and  a scalar  dilation, $\phi$,  since    the three of them complete  the bosonic   massless sector of a closed string.  Their   low energy effective action is of  the well-known  form:
\be
\dis{S_{\eff}=\int\!\rd x^{D}\sqrt{-g}e^{-2\phi}\left(R+4\partial_{\mu}\phi\partial^{\mu}\phi-\textstyle{\frac{1}{12}}H_{\lambda\mu\nu}H^{\lambda\mu\nu}\right)\,,}
\label{NSaction}
\ee
where  $R$ is the scalar curvature of the metric  and $H_{\lambda\mu\nu}$ is the three-form field strength of the two-form gauge field. Here and henceforth we consider an arbitrary spacetime dimension, $D$, without restricting ourselves  to the critical values, $10$ or $26$.  Each term in  (\ref{NSaction}) is clearly invariant under the diffeomorphism as well as   the one-form gauge symmetry, 
\be
\ba{ll}
x^{\mu}\rightarrow x^{\mu}+\delta x^{\mu}\,,~~&~~
B_{\mu\nu}\rightarrow B_{\mu\nu}+\partial_{\mu}\Lambda_{\nu} - \partial_{\nu}\Lambda_{\mu}\,.
\ea
\label{diffone}
\ee
Moreover, though not manifest, the action enjoys  T-duality which mix the three companions, $g_{\mu\nu},B_{\mu\nu},\phi$ in a nontrivial manner, first noted by Buscher~\cite{Buscher:1985kb,Buscher:1987sk,Buscher:1987qj} and further studied  in \cite{Giveon:1988tt,Tseytlin:1990nb,Tseytlin:1990va,Siegel:1993xq,Siegel:1993th,Alvarez:1994wj,Giveon:1994fu,Grana:2008yw}: If we redefine the dilaton, $\phi\rightarrow d$, and set    a $2D\times 2D$ symmetric matrix, $\cH_{AB}$ from $g_{\mu\nu}$, $B_{\mu\nu}$~\cite{Duff:1989tf},  as
\be
\ba{ll}
e^{-2d}=\sqrt{-g}e^{-2\phi}\,,~&\cH_{AB}=\!\left(\ba{cc}
g^{-1}&-g^{-1}B\\
Bg^{-1}&~g-Bg^{-1}B
\ea
\right),
\ea
\label{dcH}
\ee
%%%
%%
%%\be
%%e^{-2d}=\sqrt{-g}e^{-2\phi}\,,
%%\ee
%%and  set    a $2D\times 2D$ matrix, $\cH_{AB}$, with    $g_{\mu\nu}$ and $B_{\mu\nu}$,
%%\be
%%\cH_{AB}=\left(\ba{cc}
%%g^{\mu\nu}&-g^{\mu\kappa}B_{\kappa\sigma}\\
%%B_{\rho\kappa}g^{\kappa\nu}&~g_{\rho\sigma}-B_{\rho\kappa}g^{\kappa\lambda}B_{\lambda\sigma}
%%\ea
%%\right)\,,
%%\label{gH}
%%\ee
%%%
T-duality is conveniently  realized  by an $\ODD$ rotation which acts on the $2D$-dimensional vector  indices, 
 $A,B,\cdots,$ in a standard manner, while $d$ is taken to be an $\ODD$ singlet. The $\ODD$  group is defined by the invariance of the    constant metric of the following form,
%%%
%%$\Big(\tiny\ba{cc}{\bf{0}}&{\bf{1}}\\{\bf{1}}&{\bf{0}}\ea\Big)$, 
%%%
\be
\etaodd_{AB}:={\small{{{\left(\ba{cc}0&1\\1&0\ea\right)}}\,.}}
\label{ODDeta}
\ee
Throughout the present paper, this metric  is   being  used to  freely  raise or lower the $2D$-dimensional vector  indices.  
Indeed,  Hull and Zwiebach~\cite{Hull:2009mi,Hull:2009zb}, later with Hohm~\cite{Hohm:2010jy,Hohm:2010pp} (see also \cite{Hohm:2011dz}), managed to rewrite the effective action (\ref{NSaction}) in terms of the redefined dilaton, $d$, the $2D\times 2D$ matrix, $\cH_{AB}$, and their ordinary derivatives, such that the $\ODD$ T-duality structure  became  manifest,  yet the diffeomorphism and the one-form gauge symmetry were not any more.  In their approach,  the spacetime dimension is formally doubled from  $D$ to $2D$,  with coordinates,  $x^{\mu}\rightarrow y^{A}=(\tx_{\mu},x^{\nu})$.  The new coordinates, $\tx_{\mu}$, may  be viewed  as the canonical conjugates of the winding modes of closed strings. However,  as a  field theory counterpart to  the level matching condition in closed string theories,   it is required that 
all the  fields   as well as  all of their possible products should be  annihilated by the $\ODD$ d'Alembert operator, $\partial^{2}=\partial_{A}\partial^{A}$,
\be
\ba{ll}
\partial^{2}\Phi\equiv 0\,,~~~~&~~~~
\partial_{A}\Phi_{1}\partial^{A}\Phi_{2}\equiv 0\,.
\ea
\label{constraint}
\ee
This `{level matching constraint}' -- which we  also assume   in  this paper -- actually   means  that  the theory is not truly doubled:  there is a choice of coordinates $(\tx^{\prime},x^{\prime})$, related to the original coordinates $(\tx,x)$, by an  
$\ODD$ rotation, in which all the  fields do not depend on the $\tx^{\prime}$ coordinates~\cite{Hohm:2010jy}.  Henceforth, the equivalence symbol, `$\equiv$', means an equality up to the   constraint (\ref{constraint}), or simply up to the winding coordinate independency,  \textit{i.e.} $\frac{\partial~~}{\partial\tx_{\mu}}\equiv0$.\\
\indent Combining  the two types of  the parameters,
\[
X^{A}=(\Lambda_{\mu},\delta x^{\nu})\,,
\]
the diffeomorphism and the  one-form gauge transformations (\ref{diffone}) can be  expressed in a unified fashion, 
%%%
%%~\footnote{In our convention,  $T_{[AB]}=\half(T_{AB}-T_{BA})$, \textit{etc.}}
%%%
\be
\ba{l}
\delta_{X}\cH_{AB}\equiv X^{C}\partial_{C}\cH_{AB}+2\partial_{[A}X_{C]}\cH^{C}{}_{B}+2
\partial_{[B}X_{C]}\cH_{A}{}^{C}\,,\\
\delta_{X}\left(e^{-2d}\right)\equiv\partial_{A}\left(X^{A}e^{-2d}\right)\,.
\label{cHdTr}
\ea
\ee
These expressions  can be identified as the generalized Lie derivatives  whose commutator leads to  the Courant bracket~\cite{Courant,Siegel:1993th,Gualtieri:2003dx,Grana:2008yw,Hohm:2010pp}.  In fact, in our previous work~\cite{Jeon:2010rw}, starting from the observation that $\cH_{AB}$ given in (\ref{dcH}) assumes  a generic  form of a symmetric $\ODD$ element~\footnote{The expression of $\cH_{AB}$ in (\ref{dcH}) is the most general form of a $2D{\times 2D}$ matrix satisfying, $\cH_{AB}=\cH_{BA}$, $\cH_{A}{}^{B}\cH_{B}{}^{C}=\delta_{A}{}^{C}$, and that the upper left $D{\times D}$ block is non-degenerate~\cite{Jeon:2010rw}.},  we constructed a certain differential operator  which can be made compatible with the gauge transformations (\ref{cHdTr}), being characterized by a  projection:
\be
\ba{ll}
P_{AB}=P_{BA}=\half(\etaodd+\cH)_{AB}\,,&~
P_{A}{}^{B}P_{B}{}^{C}=P_{A}{}^{C}\,.
\ea
\label{prodef}
\ee
\indent In this work, generalizing  the    results of Ref.\cite{Jeon:2010rw} (see also works by Siegel~\cite{Siegel:1993xq} and  Hassan~\cite{Hassan:1994mq}),  we propose a novel  differential geometry apt for the unifying description of the  closed string  massless sector,   which  manifests   all the   relevant  structures simultaneously:
{\it{\small{\begin{itemize}
\item $\ODD$ T-duality
\item Gauge symmetry
\begin{enumerate}
\item Double-gauge symmetry
\begin{itemize}
\item Diffeomorphism
\item One-form gauge symmetry
\end{itemize}
\item Local    Lorentz  symmetry
\end{enumerate}
\end{itemize}}}}
\noindent In particular, we  reformulate the effective action~(\ref{NSaction}) into a single term, like  
\be
\dis{S_{\eff}\equiv\int\rd y^{2D}\, e^{-2d\,}\cH^{AB}S_{AB}\,.}
\label{NSaction2}
\ee
%%%%%%%%%%%%%%%%%%%%%%%%%%%%%%%%%%%%%%%%%%%%%%%%%%%%%%%%%%%%%%%%%%%%%
%%%%%%%%%%%%%%%%%%%%%%%%%%%%%%%%%%%%%%%%%%%%%%%%%%%%%%%%%%%%%%%%%%%%%
%%%%%%%%%%%%%%%%%%%%%%%%%%%%%%%%%%%%%%%%%%%%%%%%%%%%%%%%%%%%%%%%%%%%%
\section{Semi-covariant derivative} 
Employing  the main idea of \cite{Jeon:2010rw},  we start with a differential operator,  ${\DO_{C}=\partial_{C}+\Gamma_{C}}$, which   acts  on a generic quantity  carrying $\ODD$ vector indices, 
\be
\ba{ll}
\DO_{C}\Tw_{A_{1}A_{2}\cdots A_{n}}
\!:=\!&\!\partial_{C}\Tw_{A_{1}A_{2}\cdots A_{n}}-\omega\Gamma^{B}{}_{BC}\Tw_{A_{1}A_{2}\cdots A_{n}}\\
{}&\!\!\!+
\sum_{i=1}^{n}\,\Gamma_{CA_{i}}{}^{B}\Tw_{A_{1}\cdots A_{i-1}BA_{i+1}\cdots A_{n}}\,,
\ea
\label{semi-covD}
\ee
where $\omega$ denotes the given  weight of  each field, $\Tw_{A_{1}A_{2}\cdots A_{n}}$, and the connection must satisfy,
\be
\ba{ll}
\!\!\Gamma_{CAB}+\Gamma_{CBA}=0\,,~&\Gamma_{ABC}+\Gamma_{CAB}+\Gamma_{BCA}=0\,.
\ea
\label{sympropG}
\ee
The only quantity  which has a nontrivial weight  in this paper  is   $e^{-2d\,}$  having  ${\,\omega =1}$.  Thanks to the symmetric properties~(\ref{sympropG}), the ordinary derivatives in the definition of the generalized Lie derivative~\cite{Courant,Siegel:1993th,Gualtieri:2003dx,Grana:2008yw,Hohm:2010pp} can be replaced with  our differential operator to give 
\be
\ba{ll}
\hcL_{X}\Tw_{A_{1}\cdots A_{n}}\!:=\!&\!X^{B}\DO_{B}\Tw_{A_{1}\cdots A_{n}}+\omega\DO_{B}X^{B}\Tw_{A_{1}\cdots A_{n}}\\
{}& +\sum_{i=1}^{n}2\DO_{[A_{i}}X_{B]}\Tw_{A_{1}\cdots A_{i-1}}{}^{B}{}_{A_{i+1}\cdots  A_{n}}\,,
\ea
\label{tcLnabla}
\ee
since the connection terms  cancel. \\
\indent  We fix the connection by   requiring
\be
\ba{cc}
\DO_{A}P_{BC}=0\,,~~~&~~~\DO_{A}\brP_{BC}=0\,,\\
\multicolumn{2}{c}{\DO_{A}d:=\partial_{A}d+\half\Gamma^{B}{}_{BA}=0\,,}%\\
%%%
%%\cP_{CAB}{}^{DEF}\Gamma_{DEF}=0\,,~&~\bcP_{CAB}{}^{DEF}\Gamma_{DEF}=0\,,
%%%
\ea
\label{nablapro1}
\ee
where $\brP_{AB}=(\etaodd-P)_{AB}$ corresponds to  the  `anti-chiral' projection which is  complementary to the `chiral'  projection, $P_{AB}$  in (\ref{prodef}). Further,   ${\DO_{A}d}$  is defined by the relation,   
\be
\DO_{A}(e^{-2d})=-2(\DO_{A}d)e^{-2d}\,.
\ee
It follows that 
\be
\ba{ll}
\DO_{A}\etaodd_{BC}=0\,,~~&~~\DO_{A}\cH_{BC}=0\,.
\ea
\ee 
That is to say,  our differential operator thoroughly  annihilates the closed string massless sector represented  by $d$ and $\cH_{AB}$, which indicates that  we are on a right track to achieve a  unifying description of the   massless modes~\footnote{This differs from our previous work~\cite{Jeon:2010rw} where only $\cH_{AB}$ was annihilated and the dilaton was treated separately.}.\\
\indent  In terms of  $P$, $\brP$, $d$ and their derivatives,  the connection reads  explicitly (\textit{cf.~}\cite{Jeon:2010rw}),
\be
\small{
\ba{l}
\Gamma_{CAB}=2\!\left(P\partial_{C}P\brP\right)_{[AB]}
+2\!\left({{\brP}_{[A}{}^{\!D}{\brP}_{B]}{}^{\!E}}-{P_{[A}{}^{\!D}P_{B]}{}^{\!E}}\right)\!\partial_{D}P_{EC}\\
~~~~~-\textstyle{\frac{4}{D-1}}\!\left(\brP_{C[A}\brP_{B]}{}^{\!D}+P_{C[A}P_{B]}{}^{\!D}\right)\!\left(\partial_{D}d+(P\partial^{E}P\brP)_{[ED]}\right).
\ea}
\label{connectionG}
\ee
 It is worth while to note that, similar to our previous cases~\cite{Jeon:2010rw,Jeon:2011kp}, the following derivative vanishes  due to the level matching constraint (\ref{constraint}),
 \be
P_{I}{}^{A}\brP_{J}{}^{B}\Gamma^{C}{}_{AB}\partial_{C}~\equiv~ 0\,.
\label{USEFULv}
\ee
Furthermore,  if we set 
\be
\ba{l}
\cP_{CAB}{}^{DEF}:=P_{C}{}^{D}P_{[A}{}^{[E}P_{B]}{}^{F]}+\textstyle{\frac{2}{D-1}}P_{C[A}P_{B]}{}^{[E}P^{F]D}\,,\\
\bcP_{CAB}{}^{DEF}:=\brP_{C}{}^{D}\brP_{[A}{}^{[E}\brP_{B]}{}^{F]}+\textstyle{\frac{2}{D-1}}\brP_{C[A}\brP_{B]}{}^{[E}\brP^{F]D}\,,
\ea
\ee
which satisfy
\be
\ba{l}
\cP_{CABDEF}=\cP_{DEFCAB}=\cP_{C[AB]D[EF]}\,, \\
\cP_{CAB}{}^{DEF}\cP_{DEF}{}^{GHI}=\cP_{CAB}{}^{GHI}\,,\\
\cP^{A}{}_{ABDEF}=0\,,~~~~P^{AB}\cP_{ABCDEF}=0\,,~~~~\textit{etc.}
\ea
\ee
the connection (\ref{connectionG}) belongs to the kernel  of these  rank six-projectors, 
\be
\ba{cc}
\cP_{CAB}{}^{DEF}\Gamma_{DEF}=0\,,~&~\bcP_{CAB}{}^{DEF}\Gamma_{DEF}=0\,.
\ea
\label{kernel}
\ee
In fact,  the connection given in (\ref{connectionG}) is the unique solution to satisfy  (\ref{sympropG}), (\ref{nablapro1}) and (\ref{kernel}).\\
\indent Under the double-gauge  transformations~(\ref{cHdTr}), the  connection and  the derivative (\ref{semi-covD}) transform as
{\small{
\be
\ba{l}
\!\!(\delta_{X}{-\hcL_{X}})\Gamma_{CAB}\equiv2\big[(\cP{+\bcP})_{CAB}{}^{FDE}-\delta_{C}^{~F}\delta_{A}^{~D}\delta_{B}^{~E}\big]\partial_{F}\partial_{[D}X_{E]}\,,\\
\!\!(\delta_{X}{-\hcL_{X}})\DO_{C}T_{A_{1}\cdots A_{n}}\!\!\equiv\!
\sum_{i}2(\cP{+\bcP})_{CA_{i}}\!{}^{BFDE}
\partial_{F}\partial_{[D}X_{E]}T_{\cdots B\cdots}\,.
\ea
\label{noncov}
\ee}}
Hence, they  are not double-gauge covariant. We say, a tensor is double-gauge covariant if and only if its double-gauge transformation agrees  with  the   generalized Lie derivative.  Nonetheless,    the characteristic property of our  derivative,  $\DO_{A}$, is that,   combined with the projections,  it can  generate  various   $\ODD$ and  double-gauge covariant quantities, as follows:
\be
\ba{c}
P_{C}{}^{D}{\brP}_{A_{1}}{}^{B_{1}}{\brP}_{A_{2}}{}^{B_{2}}\cdots{\brP}_{A_{n}}{}^{B_{n}}
\DO_{D}T_{B_{1}B_{2}\cdots B_{n}}\,,\\
{\brP}_{C}{}^{D}P_{A_{1}}{}^{B_{1}}P_{A_{2}}{}^{B_{2}}\cdots P_{A_{n}}{}^{B_{n}}
\DO_{D}T_{B_{1}B_{2}\cdots B_{n}}\,,\\
P^{AB}{\brP}_{C_{1}}{}^{D_{1}}{\brP}_{C_{2}}{}^{D_{2}}\cdots{\brP}_{C_{n}}{}^{D_{n}}\DO_{A}T_{BD_{1}D_{2}\cdots D_{n}}\,,\\
\brP^{AB}{P}_{C_{1}}{}^{D_{1}}{P}_{C_{2}}{}^{D_{2}}\cdots{P}_{C_{n}}{}^{D_{n}}\DO_{A}T_{BD_{1}D_{2}\cdots D_{n}}\,,\\
P^{AB}{\brP}_{C_{1}}{}^{D_{1}}{\brP}_{C_{2}}{}^{D_{2}}\cdots{\brP}_{C_{n}}{}^{D_{n}}
\DO_{A}\DO_{B}T_{D_{1}D_{2}\cdots D_{n}}\,,\\
{\brP}^{AB}P_{C_{1}}{}^{D_{1}}P_{C_{2}}{}^{D_{2}}\cdots P_{C_{n}}{}^{D_{n}}
\DO_{A}\DO_{B}T_{D_{1}D_{2}\cdots D_{n}}\,.
\ea
\label{covariant}
\ee
Here, the  latter  second order derivatives  actually follow  from the recurrent applications of the former  first order derivatives.   
The index $n$ can be trivial, such that the covariant quantities   include  {$P^{AB}\DO_{A}T_{B}$ and $\brP^{AB}\DO_{A}T_{B}$}.\newline
%%%
%%Further, after contracting the  free index, `${\,\scriptstyle{C}\,}$', in the former two with an arbitrary  covariant vector,   
%%iterative  applications of the resulting derivatives may lead to  higher order covariant  derivatives~(see \cite{Jeon:2010rw} for details). 
%%% 
\indent The above result suggests  us to call the differential operator, $\DO_{A}$,  a `semi-covariant' derivative.

%%%%%%%%%%%%%%%%%%%%%%%%%%%%%%%%%%%%%%%%%%%%%%%%%%%%%%%%%%%%%%%%%%%%%
%%%%%%%%%%%%%%%%%%%%%%%%%%%%%%%%%%%%%%%%%%%%%%%%%%%%%%%%%%%%%%%%%%%%%
%%%%%%%%%%%%%%%%%%%%%%%%%%%%%%%%%%%%%%%%%%%%%%%%%%%%%%%%%%%%%%%%%%%%%
\section{ Curvatures}  
Straightforward  computation can show that,   the usual  curvature,
\be
\!\cR_{CDAB}=\partial_{A}\Gamma_{BCD}-\partial_{B}\Gamma_{ACD}+\Gamma_{AC}{}^{E}\Gamma_{BED}-\Gamma_{BC}{}^{E}\Gamma_{AED}\,,
\ee 
set by the  connection (\ref{connectionG}), is not double-gauge covariant,  yet it  satisfies 
\be
\ba{ll}
\cR_{CDAB}=\cR_{[CD][AB]}\,,~~&~~P_{C}{}^{I}\brP_{D}{}^{J}\cR_{IJAB}=0\,.
\ea
\ee
We define, as for a   key quantity   in our formalism,
\be
S_{ABCD}:=\half\left(\cR_{ABCD}+\cR_{CDAB}-\Gamma^{E}{}_{AB}\Gamma_{ECD}\right)\,,
\ee
which can be shown, by brute force computation,   to meet
\be
\ba{c}
\!\!S_{ABCD}= S_{\{ABCD\}}:=\half(S_{[AB][CD]}+S_{[CD][AB]})\,,\\
S_{A[BCD]}=0\,,\\
P_{I}^{~A}P_{J}^{~B}\brP_{K}^{~C}\brP_{L}^{~D}S_{ABCD}\equiv0\,,\\
P_{I}^{~A}\brP_{J}^{~B}P_{K}^{~C}\brP_{L}^{~D}S_{ABCD}\equiv 0\,,\\
P_{I}{}^{A}\brP_{J}{}^{C}\cH^{BD}S_{ABCD}\equiv 0\,,
\ea
\ee
and    have a connection to a commutator, 
\be
P_{I}{}^{A}\brP_{J}{}^{B}[\DO_{A},\DO_{B}]T_{C}\equiv 2P_{I}{}^{A}\brP_{J}{}^{B}S_{CDAB}T^{D}\,.
\ee
\indent Under the double-gauge  transformations~(\ref{cHdTr}), we get 
\be
(\delta_{X}-\hcL_{X})S_{ABCD}\equiv 4\DO_{\{A}\left[(\cP{+\bcP})_{BCD\}}{}^{EFG}\partial_{E}\partial_{[F}X_{G]}\right],
\ee
from which    double-gauge and $\ODD$ T-duality  covariant,  rank two-tensor as well as    scalar follow, 
\be
\ba{ll}
P_{I}{}^{A}\brP_{J}{}^{B}S_{AB}\,,~~~&~~~\cH^{AB}S_{AB}\,.
\ea
\label{dgcov}
\ee
Here we set 
\be
S_{AB}{=S_{BA}}{:=S^{C}{}_{ACB}}\,,
\ee
which turns out to be, from direct computation, traceless,
\be
{S^{A}{}_{A}}\equiv 0\,.
\ee
Especially, the  covariant scalar  constitutes  the effective action (\ref{NSaction2}) as
\be
\cH^{AB}S_{AB}\equiv R+4\Box\phi
-4\partial_{\mu}\phi\partial^{\mu}\phi-\textstyle{\frac{1}{12}}H_{\lambda\mu\nu}H^{\lambda\mu\nu}\,,
\label{NSaction3}
\ee
and this  is   consistent  with Refs.\cite{Hohm:2010pp,Jeon:2010rw}.   \\
\indent  Under  arbitrary infinitesimal transformations of the dilaton   and  the projection (of which the latter should obey,  from (\ref{prodef}), $\delta P= P\delta P\brP+\brP\delta P P\,$), 
%%%
%%\be
%%\delta P= P\delta P\brP+\brP\delta P P\,,
%%\ee  
%%%
we get
\be
\delta S_{ABCD}=\DO_{[A}\delta\Gamma_{B]CD}+\DO_{[C}\delta\Gamma_{D]AB}\,,
\label{infGStr}
\ee
where explicitly
{\small{\[
\ba{ll}
\!{\delta\Gamma}_{CAB}\!=\!\!&\!\!2P_{[A}^{~D}\brP_{B]}^{~E}\DO_{C}\delta P_{DE}+2(\brP_{[A}^{~D}\brP_{B]}^{~E}-P_{[A}^{~D}P_{B]}^{~E})\DO_{D}\delta P_{EC}\\
{}&\!\!-\textstyle{\frac{4}{D-1}}(\brP_{C[A}\brP_{B]}^{~D}+P_{C[A}P_{B]}^{~D})(\partial_{D}\delta d+P_{E[G}\DO^{G}\delta P^{E}_{~D]})\\
{}&\!-\Gamma_{FDE\,}\delta(\cP+\bcP)_{CAB}{}^{FDE}\,.
\ea\nonumber
\]}}
\noindent Now, with (\ref{infGStr}) and  $\DO_{A}d=0$, from the manipulation,
\[
\dis{\delta S_{\eff}\equiv\int\rd y^{2D}\, 2e^{-2d\,}\left(\delta P^{AB}S_{AB}-\delta d\,{\cH^{AB}S_{AB}}\right)\,,}
\]
it is very easy to rederive the equations of motion~\cite{Hohm:2010pp,Kwak:2010ew}:
\be
\ba{ll}
P_{(I}{}^{A}\brP_{J)}{}^{B}S_{AB}=0\,,~~~&~~{\cH^{AB}S_{AB}}=0\,.
\ea
\label{EOM}
\ee 
%%%%%%%%%%%%%%%%%%%%%%%%%%%%%%%%%%%%%%%%%%%%%%%%%%%%%%%%%%%%%%%%%%%%%
%%%%%%%%%%%%%%%%%%%%%%%%%%%%%%%%%%%%%%%%%%%%%%%%%%%%%%%%%%%%%%%%%%%%%
%%%%%%%%%%%%%%%%%%%%%%%%%%%%%%%%%%%%%%%%%%%%%%%%%%%%%%%%%%%%%%%%%%%%%
\section{Double-vielbein}  
An interesting fact about    $\etaodd_{AB}$ in (\ref{ODDeta}) and $\cH_{AB}$ in (\ref{dcH})  is that, they can be simultaneously diagonalized  as (\textit{c.f.~}\cite{Hohm:2010pp,Hohm:2010xe,Hohm:2011ex}),
\be 
\ba{l}
\etaodd=\left(\ba{ll}V&\brV\ea\right){\scriptsize{\left(\ba{lr}\eta^{-1}&0\,\\\,0&-\breta\ea\right)}}\left(\ba{ll}V&\brV\ea\right)^{t}\,,\\
\cH=\left(\ba{ll}V&\brV\ea\right){\scriptsize{\left(\ba{lr}\eta^{-1}&0\\\,0&~\,\breta\ea\right)}}\left(\ba{ll}V&\brV\ea\right)^{t}\,.
\ea
\label{DIAG}
\ee
Here $\eta$ and $\breta$ are two copies of the  $D$-dimensional Minkowskian metric. Both $V$ and $\brV$  are $2D{\times D}$ matrices which  we name  `double-vielbein'.  They must satisfy
\be
\ba{lll}
V=PV\,,~&~V\eta^{-1}V^{t}=P\,,~&~V^{t}\etaodd V=\eta\,,~~~V^{t}\etaodd\brV=0\,,\\
\brV=\brP\brV\,,~&~\brV\breta\,\brV^{t}=-\brP\,,~&~\brV^{t}\etaodd\brV=-\breta^{-1}\,,
\ea
\label{VBP}
\ee
and hence  they assume   the following  general form,
\be
\ba{ll}
V_{Am}=\textstyle{\frac{1}{\sqrt{2}}}\!{{\left(\ba{c} (e^{-1})_{m}{}^{\mu}\\(B+e)_{\nu m}\ea\right)}}\!\,,
&\brV_{A}{}^{{\brn}}=\textstyle{\frac{1}{\sqrt{2}}}\!\left(\ba{c} (\bre^{-1})^{{\brn}\mu}\\(B-\bre)_{\nu}{}^{{\brn}}\ea\right)\!\,.
\ea
\label{Vform}
\ee
Here  $e_{\mu}{}^{m}$ and  $\bre_{\nu}{}^{{\brn}}$ are two copies of  the $D$-dimensional   vielbein  corresponding  to  the same  spacetime metric (\textit{c.f.~}\cite{Siegel:1993xq,Hassan:1999bv,Hassan:1999mm}),   
\be
e_{\mu}{}^{m}e_{\nu m}=\bre_{\mu}{}^{{\brn}}\bre_{\nu{\brn}}=g_{\mu\nu}\,.
\ee
We set  $B_{\mu m}=B_{\mu\nu}(e^{-1})_{m}{}^{\nu}$, $B_{\mu\brn}=B_{\mu\nu}(\bre^{-1})_{{\brn}}{}^{\nu}$, \textit{etc.}   We may  identify  $(B+e)_{\mu m}$ and $(\brB-\bre)_{\nu{\brn}}$ as   two copies of the vielbein for the winding mode coordinate, $\tx_{\mu}$, as
\be
(B+e)_{\mu}{}^{m}(B+e)_{\nu m}\!=(B-\bre)_{\mu}{}^{{\brn}}(B-\bre)_{\nu{\brn}}\!=
(g-Bg^{-1}B)_{\mu\nu}\,.
\ee
\indent The whole internal symmetry  group is, from  (\ref{DIAG}), given by a direct product of  a pair of  $D$-dimensional Lorentz groups, \textit{i.e.~}$\SOD\times\oSOD$, of which the former and the latter   respectively acts on    each unbarred and barred  small Roman alphabet index,   while the $\ODD$ T-duality group acts   only  on the capital indices.  Indeed,  if we parametrize a $2D\times 2D$ skew-symmetric matrix as
\be
h_{AB}=-h_{BA}=\left(\ba{cc}\alpha^{\mu\sigma}&-(\beta^{t})^{\mu}{}_{\rho}\\ \beta_{\nu}{}^{\sigma}&\gamma_{\nu\rho}\ea\right)
=\left(\ba{cc}-\alpha^{\sigma\mu}&  -\beta_{\rho}{}^{\mu}\\ \beta_{\nu}{}^{\sigma}&-\gamma_{\rho\nu}\ea\right)\,,
\label{hpara}
\ee
the vectorial $\soDD$ transformations of the double-vielbein  (\ref{Vform}),  
\be
\ba{ll}
\hdelta_{h} V_{A}=h_{A}{}^{B}V_{B}\,,~~~~&~~~~
\hdelta_{h}\brV_{A}=h_{A}{}^{B}\brV_{B}\,,
\ea
\label{hVbrV}
\ee
give  well-defined $\soDD$ transformation rules for the pair of vielbeins and also  the two-form  field,
\be
\ba{l}
\hdelta_{h} e_{\mu m}= (\beta_{\mu}{}^{\nu}-B_{\mu\rho}\alpha^{\rho\nu}+g_{\mu\rho}\alpha^{\rho\nu})e_{\nu m}\,,\\
\hdelta_{h} \bre_{\mu{\brn}}\,= (\beta_{\mu}{}^{\nu}-B_{\mu\rho}\alpha^{\rho\nu}-g_{\mu\rho}\alpha^{\rho\nu})\bre_{\nu {\brn}}\,,\\
\hdelta_{h} B_{\mu\nu}= \gamma_{\mu\nu}-2\beta_{[\mu}{}^{\rho}B_{\nu]\rho}-g_{\mu\rho}\alpha^{\rho\sigma} g_{\sigma\nu}-B_{\mu\rho}\alpha^{\rho\sigma} B_{\sigma\nu}\,,%\\
%\hdelta_{h} B_{\mu\nu}= (\gamma+\beta B+B\beta^{t}-g\alpha g-B\alpha B)_{\mu\nu}\,,
\ea
\label{hdeltaeeB}
\ee
where we put, ${\hdelta_{h}:=\delta_{h}-y^{A}h_{A}{}^{B}\partial_{B}}$,  for our short hand notation.   
%%%
%%\be
%%{\hdelta_{h}=\delta_{h}-y^{A}h_{A}{}^{B}\partial_{B}\equiv\delta_{h}
%%-(\tx_{\mu}\alpha^{\mu\nu}+x^{\mu}\beta_{\mu}{}^{\nu})\partial_{\nu}}\,.
%%\ee
%%%
Both   $\hdelta_{h} e_{\mu m}$ and $\hdelta_{h} \bre_{\mu \brn}$, despite of a particular  sign difference therein (\ref{hdeltaeeB}),  give the same   transformation  rule for the metric, 
\be
\hdelta_{h} g_{\mu\nu}=(\beta g+g\beta^{t}-g\alpha B- B\alpha g)_{\mu\nu}\,,
\label{hgmunu}
\ee
which agrees with $\hdelta_{h}\cH_{AB}=h_{A}{}^{C}\cH_{CB}+h_{B}{}^{C}\cH_{AC}$.  In fact, the terms of the sign difference in $\hdelta_{h} e$ and $\hdelta_{h} \bre$ (\ref{hdeltaeeB}) can  be identified as  local Lorentz transformations,  such as 
$g_{\mu\rho}\alpha^{\rho\nu}e_{\nu m}=e_{\mu}{}^{n}\alpha_{nm}$,  $g_{\mu\rho}\alpha^{\rho\nu}\bre_{\nu \brm}=\bre_{\mu}{}^{{\brn}}\alpha_{{\brn} \brm}$. Hence, they do not   contribute to the variation of the metric, $\hdelta_{h} g_{\mu\nu}$  (\ref{hgmunu}), and  the $\soDD$ transformation of $(\bre^{-1}e)^{\brm}{}_{n}$ matches   local Lorentz transformations,
\be
\hdelta_{h}(\bre^{-1}e)^{\brm}{}_{n}=2\alpha^{\brm}{}_{\brk}(\bre^{-1}e)^{\brk}{}_{n}=2(\bre^{-1}e)^{\brm}{}_{k}\alpha^{k}{}_{n}\,.
\label{eecon}
\ee
\indent Furthermore, direct computation can show that, $V_{Am}$ and $\brV_{A}{}^{{\brn}}$ are double-gauge covariant vectors,  as their diffeomorphism plus one-form gauge symmetry transformations  (\ref{diffone}) coincide with their generalized Lie derivatives, 
\be
\ba{l}
\delta_{X}V_{Am}\equiv X^{B}\partial_{B}V_{Am}+2\partial_{[A}X_{B]}V^{B}{}_{m}=\hcL_{X}V_{Am}\,,\\
\delta_{X}\brV_{A}{}^{{\brn}}\equiv X^{B}\partial_{B}\brV_{A}{}^{{\brn}}+2\partial_{[A}X_{B]}\brV^{B{\brn}}=\hcL_{X}\brV_{A}{}^{{\brn}}\,.
\ea
\label{dcovV}
\ee
\indent We also define  ``twins" of the double-vielbein, by exchanging $e_{\mu m}$ and $\bre_{\mu \brm}$ in  (\ref{Vform}),
\be
\ba{ll}
U_{A\brn}:=\textstyle{\frac{1}{\sqrt{2}}}\!{{\left(\ba{c} (\bre^{-1})_{\brn}{}^{\mu}\\(B+\bre)_{\nu \brn}\ea\right)}}\!\,,
&\brU_{A}{}^{{m}}:=\textstyle{\frac{1}{\sqrt{2}}}\!\left(\ba{c} (e^{-1})^{{m}\mu}\\(B-e)_{\nu}{}^{{m}}\ea\right)\!\,.
\ea
\label{Uform}
\ee
They are conjugate  to the double-vielbein, satisfying the identical  properties   (\ref{VBP}), with  $U\leftrightarrow V$, $\brU\leftrightarrow\brV$,  and $\eta\leftrightarrow\bar{\eta}$.  Also they are   double-gauge covariant as in (\ref{dcovV}), and apparently local Lorentz covariant. But, due to the predetermined  transformation rules of the double-vielbein, (\ref{hdeltaeeB}), the  twins  are not qualified as  $\ODD$ covariant vectors like (\ref{hVbrV}).  Rather they transform as
\be
\ba{l}
\hdelta_{h} U_{A\brn}=h_{A}{}^{B}U_{B\brn}-2U_{A\brm}\alpha^{\brm}{}_{\brn}\,,\\
\hdelta_{h}\brU_{An}=h_{A}{}^{B}\brU_{Bn}+2\brU_{A m}\alpha^{m}{}_{n}\,.
\ea
\label{hUbrU}
\ee
\indent A crucial   ability of the  double-vielbein (\ref{Vform})  and its twin (\ref{Uform}) is  that,  they can pull  back  the chiral and the  anti-chiral  $2D$  indices  to the more familiar $D$-dimensional  ones, without losing any information  since it is, from (\ref{VBP}), an invertible process. We pull back the double-gauge covariant   rank two-tensor in (\ref{EOM}), to obtain
\be
\ba{ll}
2S_{AB}V^{A}{}_{m}\brV^{B}{}_{{\brn}}\equiv&{R}_{m\brn}+2D_{m}{D}_{\brn}\phi-\textstyle{\frac{1}{4}}H_{m\mu\nu}{H}_{\brn}{}^{\mu\nu}\\
{}&-(\partial^{\lambda}\phi){H}_{\lambda m\brn}+\textstyle{\frac{1}{2}}D^{\lambda}{H}_{\lambda m\brn}\,.
\ea
\label{EOM2}
\ee
Here we put  ${D_{m}=(e^{-1})_{m}{}^{\mu}D_{\mu}}$, ${D_{\brn}=(\bre^{-1})_{\brn}{}^{\mu}D_{\mu}}$, and set   $D_{\mu}$ to be  a  $D$-dimensional differential operator which is covariant  with respect to the diffeomorphism and the pair of local Lorentz groups,   such that  ${D_{\lambda}g_{\mu\nu}=0}$, ${D_{\mu}e_{\nu m}=0}$, ${D_{\mu}\bre_{\nu \brn}=0}$. With   the standard  diffeomorphism  covariant derivative, $\trd_{\mu}$, which involves the Christoffel symbol,  and    local Lorentz connections,   $\omega_{\mu mn}{=(e^{-1})_{m}{}^{\nu}\trd_{\mu}e_{\nu n}}$,  $\bromega_{\mu \brm\brn}{=(\bre^{-1})_{\brm}{}^{\nu}\trd_{\mu}\bre_{\nu \brn}}$,  we have
\be
D_{\mu}=\trd_{\mu}+\omega_{\mu}+\bromega_{\mu}\,.
\label{DD}
\ee
Similarly, if we compute  $S_{AB}V^{A}{}_{m}\brU^{B}{}_{{n}}$,  the resulting expression is essentially identical to (\ref{EOM2}),  with the removal of the  bar symbol from the subscript index,  $\brn$.  Then, as expected from (\ref{EOM}), the symmetric and the anti-symmetric parts of it  correspond to the equations of motion  of the effective action (\ref{NSaction}) for $g_{\mu\nu}$ and $B_{\mu\nu}$ respectively.

%%%%%%%%%%%%%%%%%%%%%%%%%%%%%%%%%%%%%%%%%%%%%%%%%%%%%%%%%%%%%%%%%%%%%
%%%%%%%%%%%%%%%%%%%%%%%%%%%%%%%%%%%%%%%%%%%%%%%%%%%%%%%%%%%%%%%%%%%%%
%%%%%%%%%%%%%%%%%%%%%%%%%%%%%%%%%%%%%%%%%%%%%%%%%%%%%%%%%%%%%%%%%%%%%
\section{ Gauging the internal symmetry} 
We now consider gauging the internal symmetry and introduce    a corresponding generalized  derivative, $\cD_{A}$,  which is  semi-covariant for the double-gauge symmetry and fully covariant for the pair of local Lorentz groups. Schematically,  we write
\be
\cD_{A}=\DO_{A}+\Omega_{A}+\brOmega_{A}\,,
\label{cDA}
\ee
where $\Omega_{A}$ and $\brOmega_{A}$ are  the connections for the local Lorentz symmetry, $\SOD$ and $\oSOD$, respectively. As the analyses are parallel, we first focus on a single  local Lorentz group, $\SOD$. In order to relate the connection, $\Omega_{Amn}$,  to the double-vielbein and its twin, specifically  $V_{An}$ and    $\brU_{An}$, we propose to  impose,
\be
\ba{l}
\brP_{A}{}^{B}\cD_{B}V_{Cm}=\brP_{A}{}^{B}(\DO_{B}V_{Cm}+\Omega_{Bmn}V_{C}{}^{n})=0\,,\\
P_{A}{}^{B}\cD_{B}\brU_{Cm}=P_{A}{}^{B}(\DO_{B}\brU_{Cm}+\Omega_{Bmn}\brU_{C}{}^{n})=0\,.
\ea
\label{isT}
\ee
These equations are not only local Lorentz covariant but also,  from (\ref{covariant}),   double-gauge covariant.  As a unique solution to them, the connection is obtained,
\be
\Omega_{Amn}=\brP_{A}{}^{B}V_{Cm}\DO_{B}V^{C}{}_{n}-P_{A}{}^{B}\brU_{Cm}\DO_{B}\brU^{C}{}_{n}\,.
\label{ChiralOmega}
\ee
Indeed, one can check straightforwardly    that, the right hand side  of (\ref{ChiralOmega})  is a double-gauge covariant vector,   and   that it  transforms   properly under the  local Lorentz transformation. 
Similarly for $\oSOD$, we get
\be
\ba{ll}
\brOmega_{A\brm\brn}&=\brP_{A}{}^{B}U_{C\brm}\DO_{B}U^{C}{}_{\brn}-P_{A}{}^{B}\brV_{C\brm}\DO_{B}\brV^{C}{}_{\brn}\\
{}&=\left[(\bre^{-1}e)(\Omega_{A}+\partial_{A})(e^{-1}\bre)\right]_{\brm\brn}\,.
\ea
\label{AntichiralOmega}
\ee
Upon the level matching constraint, they reduce to
\be
\ba{ll}
\Omega_{A}\equiv
\left(\ba{c}-\half H^{\mu}\\ 
\omega_{\nu}-\half B_{\nu\rho}H^{\rho}
\ea\right)\,,&
\brOmega_{A}\equiv
\left(\ba{c}-\half \brH^{\mu}\\ 
\bromega_{\nu}-\half B_{\nu\rho}\brH^{\rho}
\ea\right)\,,
\ea
\ee
where we let  $(H_{\mu})_{mn}=H_{\mu mn}$, $(\brH_{\mu})_{\brm\brn}=H_{\mu\brm\brn}$,  \textit{etc.}  \newline
\indent Now,   following  \cite{Jeon:2011kp}, with the semi-covariant derivative, $\DO_{A}$,  we define   a modified {Cartan-Maurer} curvature,
\be
\cF_{AB}:=\DO_{A}\Omega_{B}-\DO_{B}\Omega_{A}+[\Omega_{A},\Omega_{B}]\,.
\ee
After projecting its  two $\ODD$  vector indices  into opposite chiralities, we acquire a   rank four-tensor,
\be
\ba{l}
\cC_{mnpq}:=
2(\cF_{AB})_{mn}V^{A}{}_{p}\brU^{B}{}_{q}\\
{}~\equiv
R_{mnpq}+D_{(p}H_{q)mn}-\textstyle{\frac{1}{4}}H_{mn}{}^{l}H_{pql}-\textstyle{\frac{3}{4}}H_{m[n}{}^{l}H_{pq]l}\,,
\ea
\label{FOUR}
\ee
which is gauge, \textit{i.e.~}double-gauge plus  local Lorentz,  covariant.  However,  since the twin of the double-vielbein   is not an $\ODD$ covariant  vector (\ref{hUbrU}), the connection (\ref{ChiralOmega}) and consequently the  rank four-tensor (\ref{FOUR}) are not fully $\ODD$ covariant. Indeed, straightforward  computation  can show
 
\be
\hdelta_{h}\cC_{mnpq}=2\cC_{mnpr}\alpha^{r}{}_{q}-4V^{A}{}_{p}\brU^{B}{}_{q}\cD_{B}\cD_{A}\alpha_{mn}\,.
\ee
Turning off the $\ODD$ parameter, $\alpha^{\mu\nu}$, in   (\ref{hpara})  breaks  $\ODD$ to ${\mathbf{O}(D)\rtimes\GL(D)}$, as the  remaining ones, $\beta_{\mu}{}^{\nu}$ and $\gamma_{\mu\nu}$, form  $\mathbf{gl}(D)$ and 
$\mathbf{so}(D)$, respectively. Thus, the rank four-tensor (\ref{FOUR}) is not $\ODD$ but  ${\mathbf{O}(D)\rtimes\GL(D)}$  covariant. \newline
\indent Since  $\brOmega_{A}$ is linked  to $\Omega_{A}$  by a gauge transformation-like relation (\ref{AntichiralOmega}), the  corresponding rank four-tensor to  the other local  Lorentz group, $\oSOD$,   is essentially identical to (\ref{FOUR}), replacing the unbarred indices to the   barred indices, $C_{mnpq}\rightarrow\cC_{\brm\brn\brp\brq}$.\\
\indent Alternatively, we may consider gauging only the diagonal subgroup of ${\SOD\times\oSOD}$, \textit{i.e.~}a single Lorentz group  acting on both the barred and the  unbarred small Roman alphabet  indices simultaneously.  Then, the corresponding connection, $\Omega^{\prime}_{A}=\brP_{A}{}^{B}V_{C}\DO_{B}V^{C}-P_{A}{}^{B}\brV_{C}\DO_{B}\brV^{C}$,  as well as  its rank-four tensor   are  all  fully   $\ODD$ covariant. Choosing the gauge, $e_{\mu m}\equiv\bre_{\mu\brm}$,   spontaneously  breaks  the T-duality group, $\ODD$ to  ${\mathbf{O}(D)\rtimes\GL(D)}$,  and    the internal symmetry, ${\SOD\times\oSOD}$ to the  diagonal subgroup.    Further, it reduces the fully $\ODD$ covariant rank four-tensor to $\cC_{mnpq}$   in (\ref{FOUR}).  In this way,  $\cC_{mnpq}$  can be identified as  a  gauge fixed version  of the fully $\ODD$ covariant rank four-tensor.  \newline
\indent Finally, we pull back  the covariant derivatives in (\ref{covariant}),
\be
\ba{l}
V^{A}{}_{l}\cD_{A}T_{\brk_{1}\brk_{2}\cdots \brk_{n}}\equiv\textstyle{\frac{1}{\sqrt{2}}\,}\doD_{l}T_{\brk_{1}\brk_{2}\cdots \brk_{n}}\,,\\
\brV^{A}{}_{\brl}\cD_{A}T_{k_{1}k_{2}\cdots k_{n}}\equiv\textstyle{\frac{1}{\sqrt{2}}\,}\doD_{\brl}T_{k_{1}k_{2}\cdots k_{n}}\,,\\
P^{AB}\cD_{A}T_{B\brk_{1}\cdots \brk_{n}}\equiv\textstyle{\frac{1}{\sqrt{2}}\,}\doD^{l}T_{l\brk_{1}\cdots \brk_{n}}-{\scriptstyle{\sqrt{2}}\,}\doD^{l}\phi\, T_{l\brk_{1}\cdots \brk_{n}}\,,\\
\brP^{AB}\cD_{A}T_{Bk_{1}\cdots k_{n}}\equiv-\textstyle{\frac{1}{\sqrt{2}}\,}\doD^{\brl}T_{\brl k_{1}\cdots k_{n}}+{\scriptstyle{\sqrt{2}}\,}\doD^{\brl}\phi\, T_{\brl k_{1}\cdots k_{n}}\,,\\
P^{AB}\cD_{A}\cD_{B}T_{\brk_{1}\cdots \brk_{n}}\!\equiv\half\doD^{\mu}\doD_{\mu}T_{\brk_{1}\cdots \brk_{n}}-\doD^{\mu}\phi\, \doD_{\mu}T_{\brk_{1}\cdots \brk_{n}}\,,\\
\brP^{AB}\cD_{A}\cD_{B}T_{k_{1}\cdots k_{n}}\!\equiv-\half\doD^{\mu}\doD_{\mu}T_{k_{1}\cdots k_{n}}+\doD^{\mu}\phi \,\doD_{\mu}T_{k_{1}\cdots k_{n}}\,,
\ea
\ee
where we put, as for  $D$-dimensional tensors which are  $\ODD$ singlets, 
{\small{\[
T_{l_{1}\cdots l_{m}\brk_{1}\cdots \brk_{n}}\!=T_{A_{1}\cdot\cdot A_{m}B_{1}\cdot\cdot B_{n}}V^{A_{1}}{}_{l_{1}}\!{\cdots} V^{A_{m}}{}_{l_{m}}
\brV^{B_{1}}{}_{\brk_{1}}\!{\cdots} \brV^{B_{n}}{}_{\brk_{n}},
\]}}
and, instead of $D_{\mu}$ in (\ref{DD}),  we set, for  ${\doD_{m}=(e^{-1})_{m}{}^{\mu}\doD_{\mu}}$ and ${\doD_{\brn}=(\bre^{-1})_{\brn}{}^{\mu}\doD_{\mu}}$, 
\be
\doD_{\mu}:=\trd_{\mu}+(\omega_{\mu}+\half H_{\mu})+(\bromega_{\mu}-\half \brH_{\mu})\,.
\ee

%%%
%%%%%%%%%%%%%%%%%%%%%%%%%%%%%%%%%%%%%%%%%%%%%%%%%%%%%%%%%%%%%%%%%%%%%
%%%%%%%%%%%%%%%%%%%%%%%%%%%%%%%%%%%%%%%%%%%%%%%%%%%%%%%%%%%%%%%%%%%%%
%%%%%%%%%%%%%%%%%%%%%%%%%%%%%%%%%%%%%%%%%%%%%%%%%%%%%%%%%%%%%%%%%%%%%
\section{Discussion}   
Since the internal symmetry is given by  the direct product of two $D$-dimensional Lorentz groups,  the corresponding  gamma matrices  are a pair of  $D$-dimensional sets,  rather than a single $2D$-dimensional set. Hence  the size of spinors    will not  increase exponentially, from $2^{D/2}$ to $2^{D}$, but  will remain  the same. This seems to be a desired property while attempting to supersymmetrize  our formalism, towards  a unifying description of the type IIA and IIB supergravities, \textit{e.g.~}~\cite{Hassan:1994mq,Hassan:1999bv,Hassan:1999mm}.\newline
\indent Our results, especially the  rank four-tensor (\ref{FOUR}), may provide  powerful toolkits  to organize,   in  a simple fashion, the higher order  derivative corrections to the stringy effective actions where the four-index  Riemann or Weyl  tensors are  known to play a crucial role~\cite{Green:1984sg,Gross:1986iv,Grisaru:1986vi,Metsaev:1987zx,Bergshoeff:1994dg,Meissner:1996sa,Peeters:2000qj}.\newline
\indent  Application to doubled sigma models~\cite{Hull:2004in,Hull:2006va,Berman:2007xn,Berman:2007yf}, connection to generalized geometry~\cite{Hitchin:2004ut,Hitchin:2010qz}, or  generalization  to $\cM$-theory~\cite{Berman:2010is,Berman:2011pe}  are also of interest. \\
\indent  In the   stringy  differential geometry we have proposed,   the dilaton, $d$,   appears only explicitly  as the overall factor of the action, and  its derivatives are completely  absorbed into the connection (\ref{connectionG}), which therefore implies  the tight symmetric structure of our formalism. Furthermore,  it appears that  the natural ``cosmological constant term" is nothing but 
\be
\dis{\int\rd y^{2D}\, e^{-2d}\Lambda\equiv \int\rd x^{D}\,\sqrt{-g}e^{-2\phi}\Lambda}\,,
\ee
since $e^{-2d}$ is the only $\ODD$ singlet, scalar density with the weight of unity,  providing  the `volume form' for the double field theory action. As $\phi$  dynamically  grows, this term  becomes exponentially suppressed, irrespective of the choice of the frame, \textit{i.e.~}string or Einstein (see \textit{e.g.~}\cite{Halliwell:1986ja,Tseytlin:1990hn,Meissner:1991ge,Lidsey:1999mc}).   In this way, the notion of the cosmological constant  naturally changes in our stringy differential geometry. This might   provide a new spin on  the cosmological constant problem. \\
\indent It has  been said that string theory is a piece of $21$st century physics that happened to fall into the $20$th century~\footnote{{\textit{Anonymous  Italian.}}}.  Perhaps, our formalism  might  provide some  clue  to a new framework   beyond Riemannian geometry. \\

%%%%%%%%%%%%%%%%%%%%%%%%%%%%%%%%%%%%%%%%%%%%%%%%%%%%%%%%%%%
%%%%%%%%%%%%%%%%%%%%%%%%%%%%%%%%%%%%%%%%%%%%%%%%%%%%%%%%%%%%%
\section*{Acknowledgements} We wish to thank  Neil Copland and Michael Green for helpful comments.  The work was supported by the National Research Foundation of Korea\,(NRF) grants  funded by the Korea government\,(MEST) with the grant numbers,  2005-0049409 (CQUeST)  and 2010-0002980.


\begin{thebibliography}{99}

%\cite{Buscher:1985kb}
\bibitem{Buscher:1985kb}
  T.~H.~Buscher,
  %``Quantum Corrections And Extended Supersymmetry In New Sigma Models,''
  Phys.\ Lett.\  B {\bf 159} (1985) 127.
  %%CITATION = PHLTA,B159,127;%%
  
  %\cite{Buscher:1987sk}
\bibitem{Buscher:1987sk}
  T.~H.~Buscher,
  %``A Symmetry of the String Background Field Equations,''
  Phys.\ Lett.\  B {\bf 194} (1987) 59.
  %%CITATION = PHLTA,B194,59;%%
  
  
  %\cite{Buscher:1987qj}
\bibitem{Buscher:1987qj}
  T.~H.~Buscher,
  %``Path Integral Derivation of Quantum Duality in Nonlinear Sigma Models,''
  Phys.\ Lett.\  B {\bf 201} (1988) 466.
  %%CITATION = PHLTA,B201,466;%%
  
   %\cite{Giveon:1988tt}
\bibitem{Giveon:1988tt}
  A.~Giveon, E.~Rabinovici and G.~Veneziano,
  %``Duality in String Background Space,''
  Nucl.\ Phys.\  B {\bf 322} (1989) 167.
  %%CITATION = NUPHA,B322,167;%%
  
  

%\cite{Tseytlin:1990nb}
\bibitem{Tseytlin:1990nb}
  A.~A.~Tseytlin,
  %``Duality Symmetric Formulation Of String World Sheet Dynamics,''
  Phys.\ Lett.\  B {\bf 242}, 163 (1990).
  %%CITATION = PHLTA,B242,163;%%

%\cite{Tseytlin:1990va}
\bibitem{Tseytlin:1990va}
  A.~A.~Tseytlin,
  %``Duality Symmetric Closed String Theory And Interacting Chiral Scalars,''
  Nucl.\ Phys.\  B {\bf 350}, 395 (1991).
  %%CITATION = NUPHA,B350,395;%%



%\cite{Siegel:1993xq}
\bibitem{Siegel:1993xq}
  W.~Siegel,
  %``Two vierbein formalism for string inspired axionic gravity,''
  Phys.\ Rev.\  D {\bf 47}, 5453 (1993)
  [arXiv:hep-th/9302036].
  %%CITATION = PHRVA,D47,5453;%%

%\cite{Siegel:1993th}
\bibitem{Siegel:1993th}
  W.~Siegel,
  %``Superspace duality in low-energy superstrings,''
  Phys.\ Rev.\  D {\bf 48}, 2826 (1993)
  [arXiv:hep-th/9305073].
  %%CITATION = PHRVA,D48,2826;%%

%\cite{Alvarez:1994wj}
\bibitem{Alvarez:1994wj}
  E.~Alvarez, L.~Alvarez-Gaume and Y.~Lozano,
  %``A Canonical approach to duality transformations,''
  Phys.\ Lett.\  B {\bf 336} (1994) 183
  [arXiv:hep-th/9406206].
  %%CITATION = PHLTA,B336,183;%%

%\cite{Giveon:1994fu}
\bibitem{Giveon:1994fu}
  A.~Giveon, M.~Porrati and E.~Rabinovici,
  %``Target space duality in string theory,''
  Phys.\ Rept.\  {\bf 244}, 77 (1994)
  [arXiv:hep-th/9401139].
  %%CITATION = PRPLC,244,77;%%



%\cite{Grana:2008yw}
\bibitem{Grana:2008yw}
  M.~Grana, R.~Minasian, M.~Petrini and D.~Waldram,
  %``T-duality, Generalized Geometry and Non-Geometric Backgrounds,''
  JHEP {\bf 0904} (2009) 075
  [arXiv:0807.4527 [hep-th]].
  %%CITATION = JHEPA,0904,075;%%  


%\cite{Duff:1989tf}
\bibitem{Duff:1989tf}
  M.~J.~Duff,
  %``DUALITY ROTATIONS IN STRING THEORY,''
  Nucl.\ Phys.\  B {\bf 335} (1990) 610.
  %%CITATION = NUPHA,B335,610;%%





%\cite{Hull:2009mi}
\bibitem{Hull:2009mi}
  C.~Hull and B.~Zwiebach,
  %``Double Field Theory,''
  JHEP {\bf 0909}, 099 (2009)
  [arXiv:0904.4664 [hep-th]].
  %%CITATION = JHEPA,0909,099;%%

%\cite{Hull:2009zb}
\bibitem{Hull:2009zb}
  C.~Hull and B.~Zwiebach,
  %``The gauge algebra of double field theory and Courant brackets,''
  JHEP {\bf 0909}, 090 (2009)
  [arXiv:0908.1792 [hep-th]].
  %%CITATION = JHEPA,0909,090;%%

%\cite{Hohm:2010jy}
\bibitem{Hohm:2010jy}
  O.~Hohm, C.~Hull and B.~Zwiebach,
  %``Background independent action for double field theory,''
  JHEP {\bf 1007}, 016 (2010)
  [arXiv:1003.5027 [hep-th]].
  %%CITATION = JHEPA,1007,016;%%



%\cite{Hohm:2010pp}
\bibitem{Hohm:2010pp}
  O.~Hohm, C.~Hull and B.~Zwiebach,
  %``Generalized metric formulation of double field theory,''
  JHEP {\bf 1008}, 008 (2010)
  [arXiv:1006.4823 [hep-th]].
  %%CITATION = JHEPA,1008,008;%%



%\cite{Hohm:2011dz}
\bibitem{Hohm:2011dz}
  O.~Hohm,
  %``On factorizations in perturbative quantum gravity,''
  JHEP {\bf 1104}, 103 (2011)   [arXiv:1103.0032 [hep-th]].
  %%CITATION = JHEPA,1104,103;%%




%%%
%%\cite{Polchinski:1998rq}
%\bibitem{Polchinski:1998rq}
%  J.~Polchinski,
%  %``String theory. Vol. 1: An introduction to the bosonic string,''
%%\href{http://www.slac.stanford.edu/spires/find/hep/www?irn=4634799}{SPIRES entry}
%{\it  Cambridge, UK: Univ. Pr. (1998) 402 p}
%%%


%\cite{Courant}
\bibitem{Courant}
T. Courant, “Dirac Manifolds,”  Trans. Amer. Math. Soc. {\bf 319:} 631-661, 1990.
%%CITATION = TAMTA,319,631;%%



%\cite{Gualtieri:2003dx}
\bibitem{Gualtieri:2003dx}
  M.~Gualtieri, Ph.D. Thesis 
  ``Generalized complex geometry,''     arXiv:math/0401221.
  %%CITATION = MATH/0401221;%%

  

%\cite{Jeon:2010rw}
\bibitem{Jeon:2010rw}
  I.~Jeon, K.~Lee and J.-H.~Park,
  %``Differential geometry with a projection: Application to double field
  %theory,''
  JHEP {\bf 1104} (2011) 014
  [arXiv:1011.1324 [hep-th]].
  %%CITATION = JHEPA,1104,014;%%
  
  
 %\cite{Hassan:1994mq}
\bibitem{Hassan:1994mq}
  S.~F.~Hassan,
  %``O(d,d:R) deformations of complex structures and extended world sheet
  %supersymmetry,''
  Nucl.\ Phys.\  B {\bf 454} (1995) 86
  [arXiv:hep-th/9408060].
  %%CITATION = NUPHA,B454,86;%%


 
  %\cite{Jeon:2011kp}
\bibitem{Jeon:2011kp}
  I.~Jeon, K.~Lee and J.-H.~Park,
  %``Double field formulation of Yang-Mills theory,''
   Phys.\ Lett.\  {\bf B} (2011), doi:10.1016/j.physletb.2011.05.051   [arXiv:1102.0419 [hep-th]].
  %%CITATION = ARXIV:1102.0419;%%

  

%\cite{Kwak:2010ew}
\bibitem{Kwak:2010ew}
  S.~K.~Kwak,
  %``Invariances and Equations of Motion in Double Field Theory,''
  JHEP {\bf 1010} (2010) 047
  [arXiv:1008.2746 [hep-th]].
  %%CITATION = JHEPA,1010,047;%%
  


 
  
  
  %\cite{Hohm:2010xe}
\bibitem{Hohm:2010xe}
  O.~Hohm and S.~K.~Kwak,
  %``Frame-like Geometry of Double Field Theory,''
  J.\ Phys.\ A  {\bf 44} (2011) 085404
  [arXiv:1011.4101 [hep-th]].
  %%CITATION = JPAGB,A44,085404;%%
  
%\cite{Hohm:2011ex}
\bibitem{Hohm:2011ex}
  O.~Hohm and S.~K.~Kwak,
  %``Double Field Theory Formulation of Heterotic Strings,''
  arXiv:1103.2136 [hep-th].
  %%CITATION = ARXIV:1103.2136;%%
  
 

%\cite{Hassan:1999bv}  TWO Vielbein
\bibitem{Hassan:1999bv}
  S.~F.~Hassan,
  %``T duality, space-time spinors and RR fields in curved backgrounds,''
  Nucl.\ Phys.\  B {\bf 568} (2000) 145
  [arXiv:hep-th/9907152].
  %%CITATION = NUPHA,B568,145;%%
  
  %\cite{Hassan:1999mm}
\bibitem{Hassan:1999mm}
  S.~F.~Hassan,
  %``SO(d,d) transformations of Ramond-Ramond fields and space-time spinors,''
  Nucl.\ Phys.\  B {\bf 583} (2000) 431
  [arXiv:hep-th/9912236].
  %%CITATION = NUPHA,B583,431;%%
  
%%%%%% HIGHER ORDER DERIVATIVE CORRECTION %%%%%%%%%%%%%%%%%%%
%\cite{Green:1984sg}
\bibitem{Green:1984sg}
  M.~B.~Green and J.~H.~Schwarz,
  %``Anomaly Cancellation in Supersymmetric D=10 Gauge Theory and Superstring
  %Theory,''
  Phys.\ Lett.\  B {\bf 149} (1984) 117.
  %%CITATION = PHLTA,B149,117;%%

%\cite{Gross:1986iv}
\bibitem{Gross:1986iv}
  D.~J.~Gross and E.~Witten,
  %``Superstring Modifications of Einstein's Equations,''
  Nucl.\ Phys.\  B {\bf 277} (1986) 1.
  %%CITATION = NUPHA,B277,1;%%
  
%\cite{Grisaru:1986vi}
\bibitem{Grisaru:1986vi}
 M.~T.~Grisaru, D.~Zanon,
 %``Sigma Model Superstring Corrections To The Einstein-hilbert Action,''
 Phys.\ Lett.\  {\bf B177 } (1986)  347.



%\cite{Metsaev:1987zx}
\bibitem{Metsaev:1987zx}
  R.~R.~Metsaev and A.~A.~Tseytlin,
  %``Order alpha-prime (Two Loop) Equivalence of the String Equations of Motion
  %and the Sigma Model Weyl Invariance Conditions: Dependence on the Dilaton and
  %the Antisymmetric Tensor,''
  Nucl.\ Phys.\  B {\bf 293} (1987) 385.
  %%CITATION = NUPHA,B293,385;%%

%\cite{Bergshoeff:1994dg}
\bibitem{Bergshoeff:1994dg}
  E.~Bergshoeff, I.~Entrop and R.~Kallosh,
  %``Exact Duality In String Effective Action,''
  Phys.\ Rev.\  D {\bf 49} (1994) 6663
  [arXiv:hep-th/9401025].
  %%CITATION = PHRVA,D49,6663;%%


 %\cite{Meissner:1996sa}
\bibitem{Meissner:1996sa}
  K.~A.~Meissner,
  %``Symmetries of higher order string gravity actions,''
  Phys.\ Lett.\  B {\bf 392} (1997) 298
  [arXiv:hep-th/9610131].
  %%CITATION = PHLTA,B392,298;%%
  
  %\cite{Peeters:2000qj}
\bibitem{Peeters:2000qj}
  K.~Peeters, P.~Vanhove and A.~Westerberg,
  %``Supersymmetric higher derivative actions in ten-dimensions and
  %eleven-dimensions, the associated superalgebras and their formulation in
  %superspace,''
  Class.\ Quant.\ Grav.\  {\bf 18} (2001) 843
  [arXiv:hep-th/0010167].
  %%CITATION = CQGRD,18,843;%%

%%%%%%%%%%%%%%%%%





  

%\cite{Hull:2004in}
\bibitem{Hull:2004in}
  C.~M.~Hull,
  %``A geometry for non-geometric string backgrounds,''
  JHEP {\bf 0510}, 065 (2005)
  [arXiv:hep-th/0406102].
  %%CITATION = JHEPA,0510,065;%%

%\cite{Hull:2006va}
\bibitem{Hull:2006va}
  C.~M.~Hull,
  %``Doubled geometry and T-folds,''
  JHEP {\bf 0707}, 080 (2007)
  [arXiv:hep-th/0605149].
  %%CITATION = JHEPA,0707,080;%%



%\cite{Berman:2007xn}
\bibitem{Berman:2007xn}
  D.~S.~Berman, N.~B.~Copland and D.~C.~Thompson,
  %``Background Field Equations for the Duality Symmetric String,''
  Nucl.\ Phys.\  B {\bf 791} (2008) 175
  [arXiv:0708.2267 [hep-th]].
  %%CITATION = NUPHA,B791,175;%%

%\cite{Berman:2007yf}
\bibitem{Berman:2007yf}
  D.~S.~Berman and D.~C.~Thompson,
  %``Duality Symmetric Strings, Dilatons and O(d,d) Effective Actions,''
  Phys.\ Lett.\  B {\bf 662} (2008) 279
  [arXiv:0712.1121 [hep-th]].
  %%CITATION = PHLTA,B662,279;%%

%\cite{Hitchin:2004ut}
\bibitem{Hitchin:2004ut}
  N.~Hitchin,
  %``Generalized Calabi-Yau manifolds,''
  Quart.\ J.\ Math.\ Oxford Ser.\  {\bf 54}, 281 (2003)
  [arXiv:math/0209099].
  %%CITATION = QJMAA,54,281;%%

%\cite{Hitchin:2010qz}
\bibitem{Hitchin:2010qz}
  N.~Hitchin,
  %``Lectures on generalized geometry,''
  arXiv:1008.0973 [math.DG].
  %%CITATION = ARXIV:1008.0973;%%
  
  
%\cite{Berman:2010is}
\bibitem{Berman:2010is}
  D.~S.~Berman and M.~J.~Perry,
  %``Generalized Geometry and M theory,''
  arXiv:1008.1763 [hep-th].
  %%CITATION = ARXIV:1008.1763;%%


%\cite{Berman:2011pe}
\bibitem{Berman:2011pe}
  D.~S.~Berman, H.~Godazgar and M.~J.~Perry,
  %``SO(5,5) duality in M-theory and generalized geometry,''
  arXiv:1103.5733 [hep-th].
  %%CITATION = ARXIV:1103.5733;%%
  
  
  %\cite{Halliwell:1986ja}
\bibitem{Halliwell:1986ja}
  J.~J.~Halliwell,
  %``Scalar Fields in Cosmology with an Exponential Potential,''
  Phys.\ Lett.\  B {\bf 185} (1987) 341.
  %%CITATION = PHLTA,B185,341;%%
  
  
  %\cite{Tseytlin:1990hn}
\bibitem{Tseytlin:1990hn}
  A.~A.~Tseytlin,
  %``Duality symmetric string theory and the cosmological constant problem,''
  Phys.\ Rev.\ Lett.\  {\bf 66} (1991) 545.
  %%CITATION = PRLTA,66,545;%%
  
 %\cite{Meissner:1991ge}
\bibitem{Meissner:1991ge}
  K.~A.~Meissner and G.~Veneziano,
  %``Manifestly O(d,d) invariant approach to space-time dependent string
  %vacua,''
  Mod.\ Phys.\ Lett.\  A {\bf 6} (1991) 3397
  [arXiv:hep-th/9110004].
  %%CITATION = MPLAE,A6,3397;%%


  
  
%\cite{Lidsey:1999mc}
\bibitem{Lidsey:1999mc}
  J.~E.~Lidsey, D.~Wands and E.~J.~Copeland,
  %``Superstring cosmology,''
  Phys.\ Rept.\  {\bf 337} (2000) 343
  [arXiv:hep-th/9909061].
  %%CITATION = PRPLC,337,343;%%
  
  
%%%
%%\cite{ForCom}
%%\bibitem{ForCom}
%%For completeness, we also note
%%{\small{\be
%%\ba{ll}
%%\delta\Gamma_{CAB}=\!&\!\!2P_{[A}^{~D}\brP_{B]}^{~E}\DO_{C}
%%\delta P_{DE}+2(\brP_{[A}^{~D}\brP_{B]}^{~E}-P_{[A}^{~D}P_{B]}^{~E})
%%\DO_{D}\delta P_{EC}\\
%%{}&\!-\textstyle{\frac{4}{D-1}}(\brP_{C[A}\brP_{B]}^{~D}+
%%P_{C[A}P_{B]}^{~D})(\partial_{D}\delta d+P_{E[G}\DO^{G}\delta P^{E}_{~D]})\\
%%{}&\!-\Gamma_{FDE\,}\delta(\cP+\bcP)_{CAB}{}^{FDE}\,.
%%\ea\nonumber
%%\ee}}
%%%
\end{thebibliography}
\end{document}